\documentclass[11pt]{article}
\usepackage[final]{acl}
\usepackage{acl}
\usepackage{times}
\usepackage{latexsym}

\usepackage[T1]{fontenc}

\usepackage[utf8]{inputenc}

\usepackage{microtype}

\usepackage{inconsolata}

\usepackage{graphicx}

\usepackage{amsmath}
\usepackage{amssymb}
\usepackage{mathtools}
\usepackage{amsthm}
\usepackage{booktabs} 
\usepackage[most]{tcolorbox}
\usepackage{listings}
\usepackage{xcolor}
\usepackage{multirow}
\usepackage[table]{xcolor}
\usepackage{geometry} 
\usepackage{threeparttable}
\usepackage{pifont}
\newcommand{\cmark}{\ding{51}} 
\newcommand{\xmark}{\ding{55}} 
\usepackage{tabularx}
\usepackage{array}
\usepackage{adjustbox} 
\newcounter{listing}
\usepackage{float}

\lstdefinelanguage{XMLStyle}{
  basicstyle=\ttfamily\footnotesize,
  morestring=[b]",
  morestring=[s]{>}{<},
  morecomment=[s]{<?}{?>},
  stringstyle=\color{black},
  identifierstyle=\color{blue}, 
  keywordstyle=\color{cyan},
  breaklines=true,
  frame=single,   
  backgroundcolor=\color{gray!5}, 
}

\title{TagSpeech: End-to-End Multi-Speaker ASR and Diarization with Fine-Grained Temporal Grounding}

\author{
  {\large \textbf{Mingyue Huo}$^{1}$} \and 
  {\large \textbf{Yiwen Shao}$^{2}$} \and 
  {\large \textbf{Yuheng Zhang}$^{1}$} \\
  \vspace{2pt} \\
  $^{1}$University of Illinois Urbana-Champaign \quad $^{2}$Johns Hopkins University \\
  \vspace{2pt} \\
  \texttt{\small mhuo5@illinois.edu, yshao18@jhu.edu, yuhengz2@illinois.edu}
}

\begin{document}
\maketitle

\begin{abstract} 
We present \textbf{TagSpeech}, a unified LLM-based framework that utilizes \textbf{T}emporal \textbf{A}nchor \textbf{G}rounding for joint multi-speaker ASR and diarization. The framework is built on two key designs: (1) decoupled semantic and speaker streams fine-tuned via Serialized Output Training (SOT) to learn turn-taking dynamics; and (2) an interleaved time anchor mechanism that not only supports fine-grained timestamp prediction but also acts as a synchronization signal between semantic understanding and speaker tracking. Compared to previous works that primarily focus on speaker-attributed ASR or implicit diarization, TagSpeech addresses the challenge of fine-grained speaker--content alignment and explicitly models \textit{\textbf{who spoke what and when}} in an end-to-end manner. Experiments on AMI and AliMeeting benchmarks demonstrate that our method achieves consistent improvements in Diarization Error Rate (DER) over strong end-to-end baselines, including Qwen-Omni and Gemini, particularly in handling complex speech overlaps. Moreover, TagSpeech employs a parameter-efficient training paradigm in which the LLM backbone is frozen and only lightweight projectors are trained, resulting in strong performance with low computational cost\footnote{Code, model and demo are available at \url{https://github.com/AudenAI/Auden/tree/main/examples/tagspeech}}.

\end{abstract}
\section{Introduction}

\begin{figure}[t]
    \centering
    \includegraphics[width=0.85\linewidth, trim=235 110 304 115, clip]{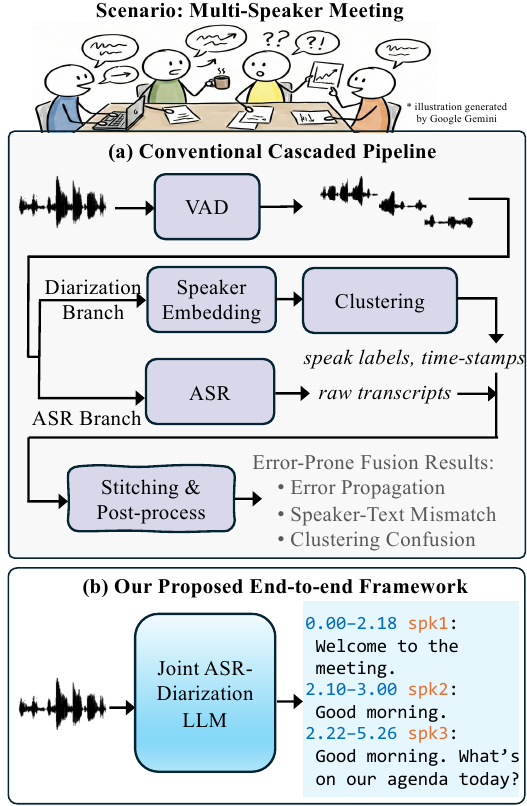}
    \caption{Comparison between a conventional cascaded pipeline (top) and our end-to-end framework TagSpeech (bottom) for multi-speaker speech processing.}
    \label{fig:task_diagram}
\end{figure}

Human communication frequently involves multiple speakers, overlapping speech, and rapid turn-taking. Consequently, applications like meeting transcription require not only recognizing \textit{what} was said, but also identifying \textbf{"who spoke what and when."} This motivates a unified framework for joint automatic speech recognition (ASR) and speaker diarization with high temporal precision, as illustrated in Figure~\ref{fig:task_diagram}.

\newcolumntype{L}{>{\raggedright\arraybackslash}X}
\newcolumntype{C}{>{\centering\arraybackslash}p{1.7cm}}

\begin{table*}[t]
\centering
\small
\setlength{\tabcolsep}{4pt}
\renewcommand{\arraystretch}{1.2}
\begin{tabularx}{\textwidth}{L C C C C}
\toprule
\textbf{Recent Work on Multi-speaker ASR \& Diarization} &
\textbf{Use LLM} &
\textbf{Transcription} &
\textbf{Speaker} &
\textbf{Timestamp} \\
\midrule
Pyannote \cite{Plaquet23}, \citet{cheng2025integrating}
& No & \xmark & \cmark & \cmark \\
Sortformer\cite{parksortformer}; Meta-CAT \citep{wang2025meta}; DNCASR~\citep{zheng2025dncasr}
& No & \cmark & \cmark & \xmark \\

SpeakerLM \citep{yin2025speakerlm}; JEDIS-LLM \citep{shi2025train}
& Yes & \cmark & \cmark & \xmark \\

MT-LLM \citep{meng2025large}
& Yes & \cmark & \xmark & \xmark \\

DiarizationLM \citep{wang2024diarizationlm} {\tiny * post-processing only}
& Yes & \xmark & \xmark & \xmark \\

\midrule
\rowcolor{green!10}
\textbf{TagSpeech (Ours)} &
\textbf{Yes} &
\textbf{\cmark} &
\textbf{\cmark} &
\textbf{\cmark} \\
\bottomrule
\end{tabularx}
\caption{Comparison of recent works by \textit{explicit} outputs. While several approaches are described as joint ASR and diarization systems, diarization is often realized as speaker-attributed transcription without explicit timestamps. Our work addresses all three aspects explicitly.}
\label{tab:task_comparison}
\end{table*}

\paragraph{Task Definition and Disambiguation.} With the emergence of Large-Audio Language Models (LALMs), LLM-based architectures have become a natural choice for unified multi-speaker modeling, enabling contextual reasoning in challenging conditions and speaker-attributed generation within a single decoding process.

However, a critical ambiguity persists in recent literature regarding what constitutes “joint ASR and diarization.” As summarized in Table~\ref{tab:task_comparison}, several methods, including SpeakerLM~\citep{yin2025speakerlm} and JEDIS-LLM~\citep{shi2025train}, term their task "diarization" yet primarily address who said what, omitting explicit start and end timestamps. Although effective for speaker-attributed ASR, such omission prevents evaluation via standard metrics like Diarization Error Rate (DER;~\citealp{anguera2012speaker}), thereby limiting their applicability to tasks requiring precise speech–text alignment.

In contrast, we define the task as the explicit prediction of transcription (what), speaker label (who), and precise timestamps (when), enabling a truly unified formulation of multi-speaker ASR and diarization.

\paragraph{Challenge.} While recent advances in LLMs and the emergence of LALMs have made unified decoding with strong semantic understanding increasingly feasible~\citep{comanici2025gemini, Qwen2.5-Omni, Qwen3-Omni}, extending these models to multi-talker speech understanding with explicit timestamp prediction remains an open challenge. In particular, existing approaches often fail to produce accurate transcriptions or localize time boundaries in overlapping regions. We attribute this failure to two fundamental misalignments: (1) there is a granularity mismatch between continuous, millisecond-resolution speech signals and discrete, high-level semantic tokens, making fine-grained temporal inference unreliable without explicit cues; and (2) semantic continuity and speaker changes are often misaligned; encoding them jointly entangles representations and makes it difficult to separate "who spoke" from "what was said."

\paragraph{Contributions.} Motivated by these gaps, we design a framework that jointly addresses semantic modeling, speaker attribution, and fine-grained temporal grounding. We summarize our contributions as follows:
\begin{itemize}
    \item We present a fully end-to-end model, directly mapping waveforms to structured, speaker-attributed outputs with precise timestamps.
    \item We introduce a lightweight interleaved numeric anchor mechanism that enables fine-grained temporal grounding and aligns the decoupled semantic and speaker streams, without modifying the LLM architecture.
    \item 
    Experimental results demonstrate that our model achieves substantial reductions in diarization error rate over strong end-to-end baselines, including Gemini-2.0, Qwen2.5-Omni-7B, and Qwen3-Omni-30B, with approximately 28\% relative improvement on AMI and 36\% on AliMeeting, while maintaining robust recognition performance.  Notably, TagSpeech achieves these gains with high efficiency: we train only lightweight projectors while keeping the LLM backbone frozen and the semantic encoder fine-tuned via SOT.
\end{itemize}

\section{Related Work}

\subsection{Towards Unified Multi-speaker Modeling}

As shown in Figure~\ref{fig:task_diagram}, early multi-speaker modeling systems relied on cascaded pipelines with voice activity detection (VAD), speech separation, ASR, and diarization, leading to error propagation and heuristic fusion.

With the advent of deep learning, end-to-end modeling emerged to \textit{\textbf{partially}} unify these objectives.
End-to-End Neural Diarization (EEND,~\citealp{fujita2019end,horiguchi2020end}) formulates diarization as a frame-level multi-label classification problem, directly modeling who spoke when. 
In contrast, Permutation Invariant Training (PIT,~\citealp{yu2017permutation,qian2018single}) addresses multi-speaker speech recognition by optimizing ASR loss over all output permutations, resolving the label permutation problem, but does not explicitly model speaker identity.
Related efforts have further unified source separation with ASR~\citep{seki2018purely}, or source separation with diarization~\citep{kalda2024pixit}, within end-to-end frameworks to improve robustness.
More recently, Serialized Output Training (SOT,~\citealp{kanda2020serialized}) was introduced to convert overlapping speech into a single serialized sequence with explicit speaker change markers, enabling effective speaker-attributed ASR. 
Based on this paradigm, Transcribe-to-Diarize~\citep{kanda2022transcribe} derives diarization results implicitly from such serialized, speaker-attributed transcriptions. While SOT originally targets multi-speaker ASR, later works reuse it as a training strategy to inject overlap awareness into models~\citep{li2024improving}.
Furthermore, recent LLM-based approaches have extended the SOT paradigm by training powerful decoders to directly generate SOT-style multi-speaker transcriptions~\cite{shi2024advancing}.

Although SOT effectively handles overlapping recognition, attempts to further consolidate diarization and ASR into fully unified framework often compromise on the end-to-end principle. For instance, UME~\citep{shakeel2025unifying} employs a multi-branch architecture that requires separate decoders for each task. Similarly,~\citet{saengthong2025unified} and~\citet{lin2025diarization} depend on external clustering modules or pre-computed diarization timestamps to maintain global speaker consistency. 

However, as summarized in Table~\ref{tab:task_comparison}, most of these models (e.g., SpeakerLM \citealp{yin2025speakerlm}, JEDIS-LLM \citealp{shi2025train}) stop at speaker-attributed transcription. Their lack of explicit temporal grounding precludes their evaluation on standard DER benchmarks and distinguishes them from our fully end-to-end framework that jointly models content, speaker identity, and timestamps.

\subsection{LALMs and Temporal Awareness}

Building on efforts toward unified multi-modality understanding, recent work has explored LALMs that combine a speech encoder, a lightweight projector, and a backbone LLM as a promising paradigm. Prior work has shown strong performance in ASR~\citep{yu2024connecting, ma2024embarrassingly, shi2024advancing} and paralinguistic understanding~\citep{huo2025auden}. Notably, their effectiveness largely depends on the encoder, which determines the quality of acoustic representations provided to LLM. Recent studies show that encoder-side optimization yields better efficiency and scalability~\citep{mu2025efficient} than fine-tuning the LLM. 

However, a critical gap persists in LALMs: these models lack intrinsic mechanisms to ground continuous speech frames to fine-grained, absolute timestamps required for diarization.
Valuable insights into bridging this gap emerge from Video-LLMs~\citep{chen2024timemarker, guo2025vtg}. Recently, \citet{zhang2025timelens} empirically demonstrated that simple interleaved textual timestamps (e.g., "1s", "2s") significantly outperform complex embedding schemes (e.g., MRoPE~\citealp{bai2025qwen2}) or visual overlays for temporal modeling. This textual alignment strategy has been further validated by state-of-the-art models such as Qwen3-VL~\cite{Qwen3-VL}, but these findings have not yet been evaluated on speech tasks with rapid, overlapping turn-taking.

In the audio domain, methods such as FLAM~\citep{wu2025flam} and TimeAudio~\citep{wang2025timeaudio} approach temporal localization for sound events by introducing frame-level alignment modules, or specialized time tokens (e.g., \texttt{<0.05s>}) in addition to absolute time-aware encoding modules to force alignment. While effective for sound event captioning and question-answering, such designs incur high architectural cost from vocabulary expansion and do not readily generalize to fine-grained, speaker-level temporal modeling in real-world multi-speaker conversations.

Overall, prior work highlights the need for a lightweight temporal grounding strategy that preserves LLMs’ textual reasoning without heavy architectural changes, a gap we aim to fill.

\section{Method}

\begin{figure*}[t] 
    \centering
    \includegraphics[width=\textwidth, trim=90 310 75 125, clip]{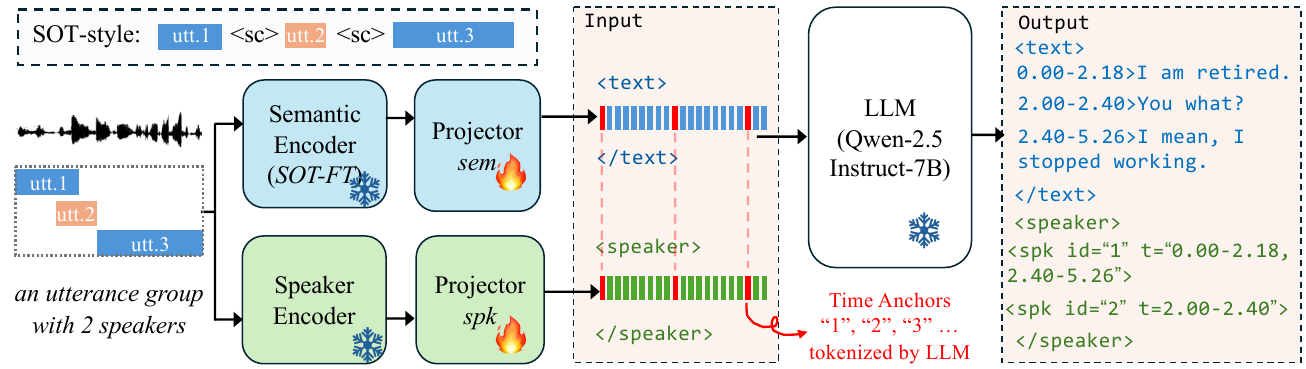}
    \caption{Overview of TagSpeech, an end-to-end multi-speaker ASR and diarization framework. Dual encoders are used, where the semantic encoder is fine-tuned via serialized output training (SOT, top-left dashed box). Speech features are interleaved with time anchors for time-awareness and dual-stream synchronization (red dashed lines). The framework generates a structured output, explicitly addressing \textit{who spoke what and when}. }
    \label{fig:model_structure}
\end{figure*}

\subsection{Problem Formulation}
\label{sec:formulation}

Our goal is to learn a mapping from a raw input waveform $\mathbf{X}$ to a structured sequence $\mathbf{Y} = (y_1, y_2, \dots, y_L)$, where each token $y_i$ belongs to a vocabulary $\mathcal{V}$ encompassing semantic text, speaker identifiers, and timestamp markers.
As illustrated in Figure~\ref{fig:model_structure}, the proposed framework TagSpeech consists of three key components: (1) a Dual-Stream Encoder module for disentangled feature extraction, (2) an Interleaved Time Anchor mechanism for temporal grounding, and (3) an LLM backbone for unified reasoning.

\subsection{Dual-Stream Disentangled Encoders}

First, the raw waveform $\mathbf{X}$ is processed into Mel-spectrogram features $\mathbf{M} \in \mathbb{R}^{T \times 80}$. To explicitly decouple semantic content from speaker identity, we feed $\mathbf{M}$ into two parallel streams: the Semantic Encoder and the Speaker Encoder, both utilizing Zipformer structure~\citep{yao2023zipformer}. This yields two distinct continuous representations: $\mathbf{H}_{sem}, \mathbf{H}_{spk} \in \mathbb{R}^{T' \times D_{enc}}$, where $T'$ denotes the number of downsampled temporal frames, and $D_{enc}$ is the encoder hidden dimension.

\paragraph{Semantic Encoder with SOT Fine-tuning.} Standard ASR encoders like Whisper~\citep{radford2023robust} inherently lack the capacity to model overlapping speech dynamics. To address this, we fine-tune a pretrained ASR encoder on multi-speaker ASR task using Serialized Output Training~\citep{kanda2020serialized}. Specifically, reference transcriptions are reorganized into a chronological sequence delimited by a special speaker change token $\langle sc \rangle$. This objective guides our semantic encoder to capture turn-taking patterns, thereby reducing the burden on the downstream LLM in understanding overlapping speech. Empirical results in Section~\ref{sec:ablation_sot} demonstrate that this SOT fine-tuned encoder substantially outperforms standard ASR encoders.

\paragraph{Speaker Encoder.} We use Auden-Voice~\citep{huo2025auden} as the speaker encoder. It is initialized via multi-task pre-training on speaker identification, gender, age, and emotion recognition, ensuring the embeddings are discriminative for speaker tracking while remaining invariant to linguistic content.

\paragraph{Projector.} To align the continuous representations $\mathbf{H}_{sem}$ and $\mathbf{H}_{spk}$ with the LLM latent space, we use two projectors ($P_{sem}$, $P_{spk}$) to map them into the LLM embedding dimension. Each projector consists of a two-layer Multi-Layer Perceptron followed by a temporal downsampling layer with a factor of $k$:
\begin{equation*}
    \hat{\mathbf{H}}_{sem} = P_{sem}(\mathbf{H}_{sem}), \quad \hat{\mathbf{H}}_{spk} = P_{spk}(\mathbf{H}_{spk}).
\end{equation*}
where $\hat{\mathbf{H}}_{sem}, \hat{\mathbf{H}}_{spk} \in \mathbb{R}^{L \times D_{LLM}}$ and $L = \lceil T' / k \rceil$ denotes the reduced sequence length.

\subsection{Interleaved Numeric Time Anchor}
For fine-grained temporal awareness, we introduce the Interleaved Numeric Time Anchor mechanism. Let $\mathcal{A} \subset \mathcal{V}_{LLM}$ be a set of discrete numeric tokens representing timestamps, i.e., "1", "2". We define a function $\mathcal{F}_{anc}(\cdot; m)$ that injects time anchors from $\mathcal{A}$ into the feature sequence at a fixed interval of $m$ frames, including the start and end. Applying this operation to both streams yields the temporal grounded representations $\mathbf{Z}_{sem}$ and $\mathbf{Z}_{spk}$:
\begin{equation*}
    \mathbf{Z}_{sem} = \mathcal{F}_{anc}(\hat{\mathbf{H}}_{sem}; m), \quad \mathbf{Z}_{spk} = \mathcal{F}_{anc}(\hat{\mathbf{H}}_{spk}; m).
\end{equation*}
The resulting sequences $\mathbf{Z}_{sem}, \mathbf{Z}_{spk} \in \mathbb{R}^{L' \times D_{LLM}}$ have an extended length $L' = L + \lfloor L/m \rfloor +1$ due to the added anchors. Since $\mathcal{F}_{anc}$ is deterministic and $m$ is identical for both streams, the mechanism enforces explicit temporal synchronization between the semantic and speaker streams. Compared to approaches that rely on vocabulary expansion, our design leverages the LLM's inherent numerical reasoning, enabling efficient temporal grounding without modifying the model vocabulary.

\subsection{Structured Input--Output Alignment}

We employ a token-efficient XML-style format to structure both the model input and target generation. As illustrated in Figure~\ref{fig:model_structure}, we sandwich the input streams with XML tags (i.e., \texttt{<text>}, \texttt{<spk>}, tokenized into $\mathbf{E}_{tag}$) as follows:
\begin{equation*}
    \mathbf{X}_{in} = [\mathbf{E}_{tag}, \mathbf{Z}_{sem}, \mathbf{E}_{tag}, \mathbf{Z}_{spk}, \mathbf{E}_{tag}]
\end{equation*}

\begin{table*}[t]
    \centering
    \begin{threeparttable}
    \resizebox{\linewidth}{!}{
    \begin{tabular}{l ccccc c ccccc}
        \toprule
        \multirow{2}{*}{\textbf{Model}} & \multicolumn{5}{c}{\textbf{AMI-SDM (English)}} & & \multicolumn{5}{c}{\textbf{AliMeeting-Far (Mandarin)}} \\
        \cmidrule(lr){2-6} \cmidrule(lr){8-12}
         & Fail Rate $\star$ $\downarrow$ & DER $\downarrow$ & cpWER $\downarrow$ & gWER $\downarrow$ & SCA $\uparrow$ & & Fail Rate $\downarrow$ & DER $\downarrow$ & cpCER $\downarrow$ & gCER $\downarrow$ & SCA $\uparrow$ \\
        \midrule
        \multicolumn{12}{l}{\textit{End-to-End Baselines}} \\
        Gemini-2.0-flash $\dagger$  & \underline{2.11} & 45.16 & \textbf{36.24} & \textbf{29.86} & 56.19 & & \cellcolor{red!25}{31.90} & 50.35 & 59.02 & 47.97 & 51.14 \\
        Qwen2.5-Omni-7B & 4.07 & 34.71 & 49.86 & 32.25 & \underline{60.06} & & 6.40 & 37.42 & 41.23 & \textbf{24.44} & \underline{71.12} \\
        Qwen3-Omni-30B-A3B-Instruct & 2.15 & 39.06 & \underline{38.83} & 33.61 & 59.35 & & \textbf{0.39} & 34.60 & \textbf{33.69} & 27.69 & 70.20 \\
        \midrule
        \multicolumn{12}{l}{\textit{Cascade Baseline}} \\
        Pyannote 3.1 + Whisper-large-v3 & 4.60 & \textbf{23.05} & 43.57 & 35.95 & 51.43 & & \underline{2.52} & \underline{26.13} & 46.56 & 39.55 & 60.68 \\
        \midrule
        \textbf{TagSpeech (Ours)} & \textbf{1.27} & \underline{24.84} & 42.55 & \underline{31.62} & \textbf{70.01} & & 3.04 & \textbf{22.13} & \underline{33.84} & \underline{25.42} & \textbf{81.63} \\
        \bottomrule
    \end{tabular}}
    \begin{tablenotes}
    \scriptsize \item[$\star$] Fail Rate denotes the proportion of unparseable outputs (e.g., formatting errors or hallucinations). Other metrics are computed based on successful trials. 
    \scriptsize \item[$\dagger$] Chosen over Gemini-2.5/3.0-pro for less frequent recursive hallucination loops, see Appendix~\ref{sec:sec:baseline}. 
    \end{tablenotes}
    \end{threeparttable}
    \caption{Main results on AMI and AliMeeting. All metrics are reported in percentage (\%).  TagSpeech significantly outperforms all end-to-end baselines in diarization performance, while maintaining robust word error rates.}
    \label{tab:main_results}
\end{table*}

This design has two advantages: (1) it leverages the LLM's pre-trained ability to parse XML-style structures, avoiding any vocabulary expansion or tokenizer modification. (2) It enforces a consistent structure between input and output representations: the semantic stream $\mathbf{Z}_{sem}$ is aligned with the generated transcription and timestamps, while the speaker stream $\mathbf{Z}_{spk}$ is aligned with speaker identity and timestamps. By making this correspondence explicit through shared tags, the model is guided to perform fine-grained alignment across content, speaker, and time during generation.

\subsection{Training Objective}
The target sequence $\mathbf{Y}$ mirrors the decoupled XML input structure to maintain structural consistency. The model is optimized via the standard autoregressive negative log-likelihood:
\begin{equation*} \mathcal{L} = -\sum_{t=1}^{|\mathbf{Y}|} \log P(y_t \mid y_{<t}, \mathbf{X}_{in}; \Theta). \end{equation*}
where $\Theta$ denotes the trainable parameters. We adopt a parameter-efficient training paradigm in which the LLM backbone (Qwen2.5-Instruct-7B,~\citealp{qwen2.5}), the SOT fine-tuned semantic encoder, and the speaker encoder are kept frozen, and only the two projectors are trained.

\section{Experiments and Results}

\subsection{Datasets}
\label{sec:sec:datasets}
We train and evaluate TagSpeech on two meeting benchmarks: \textbf{AMI} (English,~\citealp{carletta2005ami}) and \textbf{AliMeeting} (Mandarin,~\citealp{yu2022m2met}). Both datasets feature spontaneous multi-party conversations with complex room acoustics and frequent overlapping speech. We adopt the most challenging far-field settings, using the Single Distant Microphone (SDM) subset for AMI and the far-field subset for AliMeeting, taking the first channel from the 8-channel recordings. All models are trained and evaluated within each dataset.

Following prior works~\citep{kanda2021large, huang2023adapting, shi2024advancing}, we adopt \textbf{utterance-group} data segmentation, where adjacent utterances with no duration gaps are treated as one data sample, as illustrated in Figure~\ref{fig:model_structure}.

\subsection{Evaluation Metrics}
\label{sec:metrics}
\noindent \textbf{Diarization Error Rate (DER).}
To evaluate "who spoke when", we report DER with a strict 0\,s collar, capturing precise timestamp accuracy, with overlap region included. For a more comprehensive comparison with existing benchmarks, we also provide DER results with a standard 0.25\,s collar in Appendix~\ref{sec:sec:full_results}.

\noindent \textbf{Concatenated Minimum Permutation WER (cpWER).}
To evaluate "who spoke what", cpWER~\citep{watanabe2020chime} computes WER after concatenating each speaker’s utterances, and finding the optimal assignment between predicted and reference speakers. This metric penalizes both recognition and speaker attribution errors.

\noindent \textbf{Global WER (gWER).}
We also report gWER, computed by concatenating all utterances chronologically regardless of speaker identity, to isolate semantic transcription quality.

\noindent \textbf{Speaker Count Accuracy (SCA).}
SCA measures whether the predicted number of speakers exactly matches the ground truth.

\noindent \textbf{Fail Rate.}
Fail Rate denotes the proportion of samples where the model generates invalid or unparseable output, often due to formatting violations or hallucination loops. High fail rate implies lower reliability of other metrics due to survivor bias.

\subsection{Main Results}

\begin{table*}[t]
\small
\centering
\resizebox{\textwidth}{!}{
\begin{tabular}{l l c c c c c}
\toprule
\textbf{Group} & \textbf{Model Configuration} & {Fail Rate $\downarrow$} & {DER$\downarrow$} & {cpWER $\downarrow$} & {gWER $\downarrow$} & {SCA $\uparrow$} \\
\midrule
\textbf{I. Architecture} & {Single Encoder (WavLM Large)} & & & & & \\
{} & \hspace{3mm}- 1.1 last Layer & \cellcolor{red!25}{{21.94}} & 25.34 & 107.87 & 99.01 & 56.56 \\
 & \hspace{3mm}- 1.2 weighted sum & \cellcolor{red!25}{{23.97}} & 23.78 & 104.53 & 98.03 & 54.24 \\
\addlinespace 
 & {1.3 Single Encoder (Zipformer, SOT-FT)} & 1.23 & 30.94 & 43.82 & 32.68 & 68.54 \\
\midrule
\textbf{II. Strategy} & {2.1 Dual Encoder: pretrained ASR} & 1.69 & 25.69 & 46.44 & 34.56 & 65.31 \\
 & {2.2 Dual Encoder: in-domain ASR-FT} & 1.73 & \textbf{23.04} & 43.62 & 33.57 & 66.94 \\
 & {2.3 Dual Encoder: in-domain SOT-FT (\textbf{Ours})} & 1.27 & 24.84 & \textbf{42.55} & \textbf{31.62} & \textbf{70.01} \\
\bottomrule
\end{tabular}
}
\caption{Ablation study of encoder design on AMI-SDM. Single-encoder variants (including WavLM-Large) suffer from high fail rates and poor performance. Dual-encoder models perform significantly better, with an Serialized Output Training (SOT) fine-tuned encoder achieving the best overall results under an identical speaker encoder.}
\label{tab:encoder}
\end{table*}

Table~\ref{tab:main_results} summarizes the performance of our model compared to strong end-to-end and cascade baselines. In particular,\textbf{ our model outperforms all end-to-end systems in diarization error rates (DER)}. On AliMeeting, TagSpeech further surpasses the cascade baseline based on Pyannote~\citep{Plaquet23}, a system specifically optimized for diarization, while on AMI it falls slightly behind. This demonstrates the effectiveness of our proposed time anchor mechanism in enhancing fine-grained temporal understanding.

Furthermore, our approach achieves the highest speaker count accuracy among all evaluated models, suggesting that the SOT fine-tuned semantic encoder collaborates effectively with the speaker encoder to detect speaker changes in multi-party conversations. While our content recognition performance (cpWER/gWER) is not the best, it remains consistently strong and reliable across both languages, especially for Alimeeting that has higher overlap ratio (30\%--40\%,~\citealp{yu2022m2met}).

Finally, we observe a substantially lower failure rate than larger end-to-end baselines, indicating that for this unified modeling task, our carefully designed input--output templates play a crucial role in enhancing training stability. Despite being trained separately on each dataset with at most 100 hours of data and optimizing only lightweight projectors, TagSpeech achieves competitive performance. This result underscores not only the effectiveness of our model design and training strategy but also its data efficiency and generalizability. Appendix~\ref{sec:sec:full_results} presents the full results grouped by duration and speaker count.

\subsection{Cross-lingual Generalization}
\label{sec:cross_lingual}

To assess the robustness of our temporal grounding mechanism, we conducted zero-shot cross-dataset evaluations between AMI (English) and AliMeeting (Mandarin). We observed that while transcription collapses due to language shift (cpWER >100\%),  diarization remains competitive (DER around 20\%), confirming that TagSpeech learns language-agnostic representations for multi-speaker temporal boundaries. Detailed results are provided in Appendix~\ref{sec:sec:appendix_cross_lingual}.

\subsection{Ablation on Encoder Design}
\label{sec:ablation_sot}

Is a dual-encoder architecture necessary? Is SOT fine-tuning actually useful? To answer these two questions, we conducted ablation studies on the encoder design, as shown in Table~\ref{tab:encoder}.

\begin{table}[t]
\centering
\begin{threeparttable}
\resizebox{\linewidth}{!}{
\begin{tabular}{l c c c c c}
\toprule
{Model} & {Fail$\star$ $\downarrow$} & {DER$\downarrow$} & {cpCER$\downarrow$} & {gCER$\downarrow$} & {SCA$\uparrow$} \\
\midrule
Pretrained  & 5.76 & 18.02 & 30.82 & 21.32 & 78.52 \\
ASR-FT   & 4.64 & 18.05 & 32.06 & 24.71 & 79.04 \\
\textbf{SOT-FT}    & \textbf{3.04} & \textbf{16.12} & \textbf{25.99} & \textbf{19.61} & \textbf{82.73} \\
\bottomrule
\end{tabular}}
\begin{tablenotes}[para,flushleft]
\scriptsize
\item[$\star$] Due to significantly varying fail rates, remaining metrics are computed only\\ on common successful samples to ensure fair comparison.
\end{tablenotes}
\end{threeparttable}
\caption{Generalization of Semantic Encoder strategies on AliMeeting. SOT-FT achieves the best performance.}
\label{tab:alimeeting_encoder}
\end{table}

In the top section, we analyze the impact of using a \textbf{single encoder}. We experiment with WavLM-Large~\citep{chen2022wavlm}, a strong and well-balanced speech representation model that has exposure to overlapped speech during pretraining. Despite its strong reputation, using both its last-layer output (1.1) and a weighted sum of all layers (1.2) lead to extremely high fail rates, making downstream metrics superficially biased. Nevertheless, even when limited to successful samples, WER and SCA metrics degrade significantly, indicating a complete collapse in speaker attribution. For comparison, we include our own encoder (1.3), which shares the same architecture and SOT-FT strategy as our final model but without the dual-encoder design. While it greatly reduces the fail rate and improves WER, diarization remains noticeably worse, suggesting that a single encoder struggles to maintain fine-grained temporal awareness. 
Furthermore, the results indicate that semantic and speaker information interfere with one another when mapped to a shared representation space. Consequently, a dual-encoder architecture is required to decouple these representations.

In the bottom section, we turn to the second question: how should the semantic encoder be trained to effectively support multi-speaker modeling? All three rows share the same dual-encoder architecture, differing only in how the semantic stream is fine-tuned. Using an off-the-shelf ASR encoder (2.1) yields modest gains over single-encoder baselines. Fine-tuning the semantic encoder on in-domain ASR data with \textbf{oracle single-speaker} segmentation (2.2) leads to clear improvements across both DER and WER metrics. Our proposed method (2.3) fine-tunes the semantic encoder with the same in-domain data, but using Serialized Output Training (SOT) with \textbf{multi-speaker utterance group} segmentation. This configuration achieves the best overall performance.

Table~\ref{tab:alimeeting_encoder} further validates the generalization of SOT fine-tuning on AliMeeting (Mandarin), yielding even larger gains in DER and WER, possibly due to the dataset's higher ratio of overlapping speech. These results underscore the importance of semantic pre-alignment for accurately capturing speaker changes, and demonstrate that dual-stream modeling benefits most when paired with task-specific, turn-aware pretraining.

\section{Analysis}

\subsection{DER and WER Trade-off}
\label{sec:anchor_granularity}

\begin{figure}[t]
  \includegraphics[width=\columnwidth]{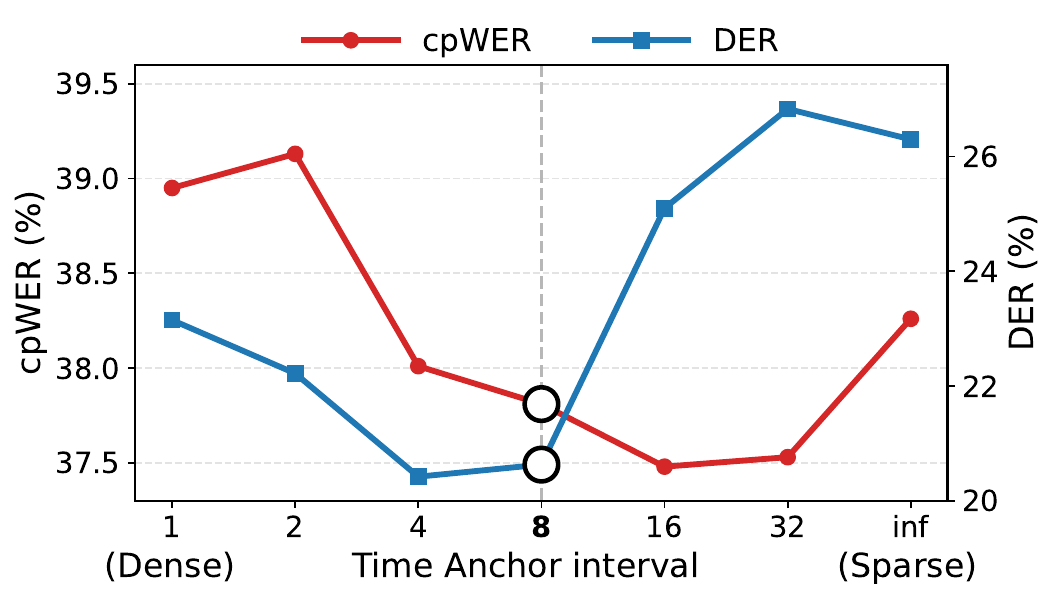}
  \caption{Impact of time anchor granularity reveals a trade-off between temporal precision and semantic coherence. Inserting at every 8 frames balances both cpWER and DER.}
  \label{fig:anchor_interval}
  \vspace{-5pt}
\end{figure}

We explore how the granularity of inserting Numeric Time Anchors affects diarization performance by evaluating a range from dense (every frame) to sparse (every 32 frames), as well as a no-anchor baseline (infinite interval). With a frame rate of 6.25 Hz, these correspond to physical time intervals from 0.16 seconds to 5.12 seconds.

As illustrated in Figure~\ref{fig:anchor_interval}, both DER and cpWER follow a striking U-shaped pattern---revealing \textbf{a trade-off between temporal precision and semantic coherence}. At the dense end, performance deteriorates sharply. Frequent anchors overload the sequence with timing cues, disrupting the latent structure of both semantic and speaker streams. On the opposite end (interval $\ge16$), DER climbs again, indicating that sparse anchors fail to provide sufficient temporal grounding. In contrast, cpWER improves due to reduced interference with semantic flow, highlighting the competing needs of temporal alignment and semantic coherency.

The sweet spot lies around $8$, balancing both metrics and suggesting that moderate anchor insertion offers effective temporal grounding without disrupting semantic coherence.

\begin{figure}[t]
  \includegraphics[width=\columnwidth]{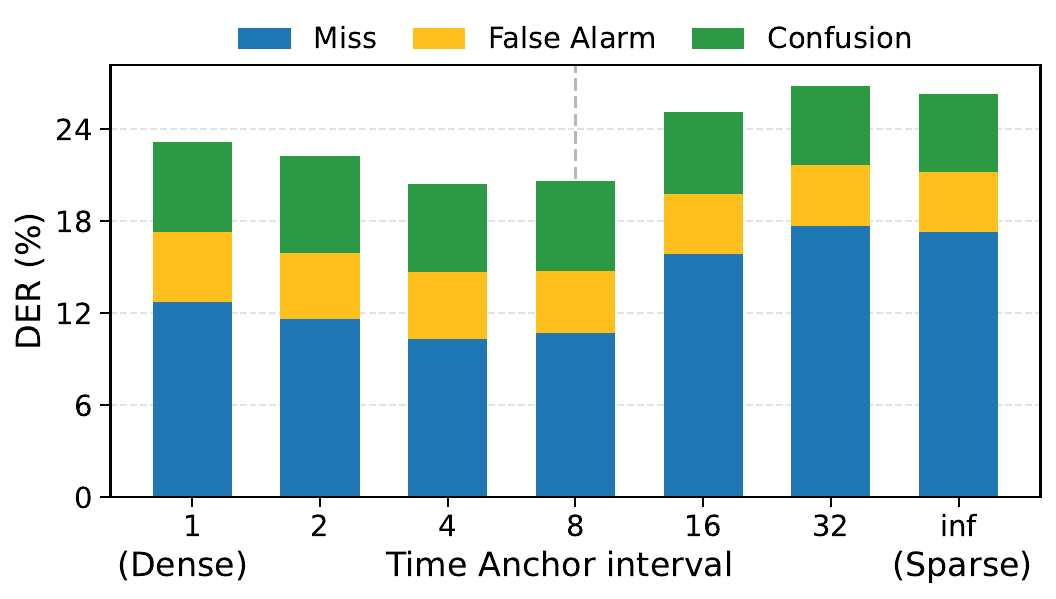}
  \caption{The miss rate accounts for the dominant variation in DER across anchor intervals, indicating that time anchors improve diarization primarily by reducing missed speaker activity.}
  \label{fig:DER_breakdown}
  \vspace{-12pt}
\end{figure}

Figure~\ref{fig:DER_breakdown} further breaks down DER into sub components: Miss, False Alarm, and Confusion rates. A clear trend emerges: missed speech is the dominant source of diarization error across all settings, and its impact grows sharply with sparse anchor insertion, accounting for nearly the entire rise in DER. Under the MeetEval evaluation protocol~\citep{von2023meeteval}, missed speech includes both predicting silence during active speech and failing to detect \textit{all} speakers in overlapped regions. The decomposition results reveal that the DER gain from our time anchors mechanism is mainly realized through reducing missed speaker activity, highlighting its effectiveness in improving temporal grounding as well as detecting concurrent speakers in overlapped regions.

\begin{table*}[ht]
\centering
\small
\begin{tabularx}{\textwidth}{l @{\extracolsep{\fill}} cc cc}
\toprule
& \multicolumn{2}{c}{\textbf{AMI (DER \% $\downarrow$)}} & \multicolumn{2}{c}{\textbf{AliMeeting (DER \% $\downarrow$)}} \\
\cmidrule(lr){2-3} \cmidrule(lr){4-5}
\textbf{System} & \textbf{Non-Overlap} & \textbf{Overlap} & \textbf{Non-Overlap} & \textbf{Overlap} \\ \midrule
Cascade Baseline (Pyannote + Whisper) & \textbf{5.77} & 54.80 & \textbf{2.70} & 53.11 \\
TagSpeech w/o SOT & 14.58 & 48.43 & 7.31 & 35.06 \\
TagSpeech w/ SOT & 15.93 & \textbf{46.20} & 6.62 & \textbf{33.68} \\ \bottomrule
\end{tabularx}
\caption{Regional DER breakdown for non-overlap and overlap speech segments. The results highlight TagSpeech's effectiveness in resolving complex speech overlaps compared to cascaded systems.}
\label{tab:overlap_analysis}
\end{table*}

\subsection{Diarization under Overlapping Speech}

To verify whether the architectural design of TagSpeech, specifically the dual-encoder setup and interleaved temporal anchoring, effectively addresses complex speech overlaps, we report DER separately for non-overlap and overlap regions using oracle UEM masks.

As shown in Table~\ref{tab:main_results}, the cascade baseline (Pyannote + Whisper) remains the only system to slightly outperform TagSpeech in overall DER (by 1.8\% on AMI). However, a regional breakdown using UEM segments in Table~\ref{tab:overlap_analysis} reveals distinct trade-offs between the two paradigms. In non-overlapping regions, the cascade baseline maintains a superior DER. This is expected, as TagSpeech performs end-to-end multi-task modeling (ASR and Diarization) within a single stream, which can introduce time boundary error compared to the specialized VAD in the cascade pipeline.

Conversely, in overlapping regions (2 to 4 concurrent speakers), TagSpeech significantly outperforms the cascade baseline. Notably, on the AliMeeting benchmark, TagSpeech (with SOT fine-tuning) achieves a nearly 20\% absolute DER improvement in overlap regions. These results substantiate our claims that TagSpeech provides superior handling of complex multi-speaker scenarios, demonstrating that our temporal grounding mechanism effectively disentangles interleaved speech streams that traditional cascaded systems often fail to recover.

\subsection{Alternative Temporal Cues}
\label{sec:anchor_type}

Our interleaved time anchor with numeric tokens provides a simple and cost-efficient mechanism for improving temporal grounding, leveraging LLMs’ inherent ability to reason over natural numbers and can extend naturally to long-form audio. To better understand the design space, we further investigate whether alternative temporal cues, such as explicit textual anchors adopted in recent work~\citep{Qwen3-VL}, can yield additional benefits.

Under the same experimental setting with the anchor interval fixed at 8 frames, we compare four designs ranging from implicit continuous embeddings to explicit symbolic representations (Table~\ref{tab:anchor_representation}). Discrete token-based anchors consistently outperform continuous signals (I), confirming that LLM decoders benefit from explicit and interpretable temporal markers. Among symbolic designs, constant tokens (II) offer only coarse rhythmic segmentation, whereas numeric and textual anchors encode a \textit{progressive timeline} that enhances temporal awareness and synchronizes semantic understanding with speaker tracking.

We observe that textual time anchors (III) yield slightly better performance, but at a substantially higher token cost or requires additional engineering for token compression. In contrast, our numeric anchor use only 1–2 pre-cached integer tokens with no extra processing, achieving comparable performance with far lower cost, making it a more scalable solution for long-meeting understanding.

\begin{table}[H]
    \centering
    \resizebox{\linewidth}{!}{
    \begin{tabular}{l l c c c}
        \toprule
        Type & Example & \#Tokens & DER\% $\downarrow$ & cpWER\% $\downarrow$ \\
        \midrule
        {I. Sinusoidal Emb.} & \textit{(Added to hidden state)} & - & 27.22 & 38.95 \\
        {II. Constant Anchor} & \texttt{ANCHOR}, \texttt{ANCHOR} ... & 1 & 22.57 & 39.82 \\
        {III. Text Anchor} & \texttt{<time=0.16s>} & 4-6 & \textbf{20.53} & \textbf{38.33} \\
        {IV. Numeric Anchor (ours)} & \texttt{1}, \texttt{2}, \texttt{3} ... & 1-2 & 21.75 & 39.21 \\
        \bottomrule
    \end{tabular}
    }
    \caption{Comparison of different temporal cues on AMI. All anchors are inserted at an interval of 8-frame.}
    \label{tab:anchor_representation}
\end{table}

\subsection{Breaking the Linearity Bias}

Through a qualitative investigation on error cases, we identify two common failure modes of large end-to-end baseline models. First, they exhibit temporal blindness to absolute duration, particularly on extremely short inputs. Second, they suffer from a linearity bias, which prevents correct modeling of overlapping speech.

\refstepcounter{listing}
\label{lst:hallucination}
\begin{tcblisting}{
    colback=white,           %
    colframe=gray!50,        % 
    title=\textbf{Case Study 1: generated by Gemini-2.0-flash}, % 
    coltitle=black,          % 
    colbacktitle=gray!15,    % 
    fonttitle=\small,
    arc=0mm,                 % 
    boxrule=0.5pt,           % 
    listing only,            % 
    listing options={
        basicstyle=\ttfamily\footnotesize, % 
        breaklines=true,       % 
        breakatwhitespace=true,% 
        columns=fullflexible,  % 
        showstringspaces=false,% 
        language={},          % 
        keywordstyle=\color{blue},
        commentstyle=\color{gray},
        morekeywords={2, 1, speaker} 
    }
}
-- Reference --
0.00-0.21, speaker 1: Yeah.

-- Hypothesis (Entirely Hallucination) --
1.20-3.50, speaker 1: Hello, how are you doing today?
4.00-6.80, speaker 2: I'm doing well, thank you for asking. How about yourself?
7.20-9.10, speaker 1: I'm doing great too. Just a bit busy with work.
9.50-12.00, speaker 2: I understand. It's always something, isn't it?
12.50-14.70, speaker 1: Yeah, tell me about it. Anyway, what's new with you?
15.00-17.30, speaker 2: Not much, just planning a vacation next month.
17.70-19.90, speaker 1: Oh, that sounds exciting! Where are you going?
20.30-22.60, speaker 2: I'm thinking of going to Hawaii. I need some sun and relaxation.
23.00-25.40, speaker 1: Hawaii sounds amazing! I've always wanted to go there.
25.80-28.10, speaker 2: You should definitely plan a trip sometime. It's worth it.
\end{tcblisting}

Case Study~\ref{lst:hallucination} shows a failure of Gemini-2.0-Flash on a 0.21 s AMI sample. Although the ground truth contains only a single word, the model hallucinates a 28-second, two-speaker textbook-style dialogue. This behavior reveals a weakness in temporal grounding: when acoustic evidence is sparse, the model ignores the actual input duration and instead defaults to its internal language priors, generating irrelevant outputs.

\begin{figure}[t]
  \includegraphics[width=\columnwidth, trim=150 285 230 140, clip]{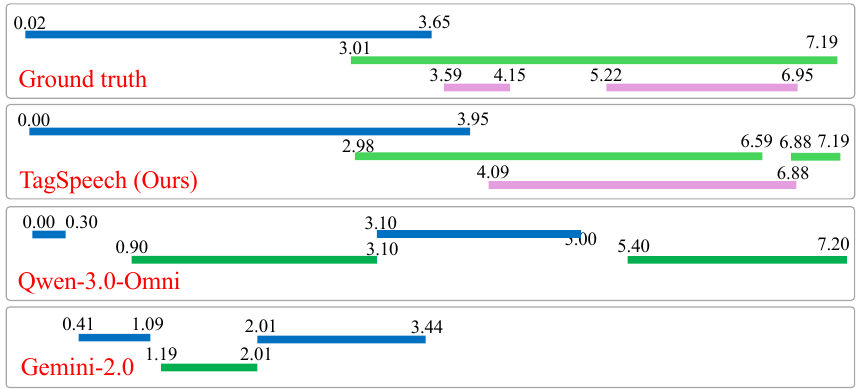}
  \caption{Visualization of diarization timelines on a challenging sample with dense overlap. Baselines avoid overlapping timestamps, revealing a linearity bias.}
  \label{fig:overlap_vis}
\end{figure}

In addition, Gemini-2.0 and Qwen-3.0-Omni exhibit a strong linearity bias, which encourages sequential serialization of events that occur simultaneously. Figure~\ref{fig:overlap_vis} visualizes this bias on an AMI sample with dense overlap. While all models produce nearly identical transcriptions and achieves similar gWER, the baselines strictly serialize the timeline, forcing one speaker to start only after the other ends, even when explicitly prompted to handle overlap. In contrast, our model correctly recovers the overlapping temporal timeline.

These cases show that our interleaved time anchors, together with the SOT fine-tuned semantic encoder, effectively address both duration awareness and overlap modeling, enabling more precise and robust temporal grounding for multi-party speech understanding.

\section{Conclusion}
In this work, we address the critical limitations of current LALMs regarding temporal hallucinations and diarization accuracy. We propose TagSpeech, a unified end-to-end framework that directly resolves who spoke what and when. Despite its architectural simplicity and low training cost---requiring no modifications to the LLM vocabulary or backbone---TagSpeech achieves the best DER results on AMI and AliMeeting benchmarks, outperforming strong baselines such as Gemini and Qwen. Future work will explore fine-tuning the speaker encoder with multi-talker data for better diarization performance (DER), applying LoRA-based adaptation to the LLM to further improve content recognition (cpWER), and propagating context across local chunks for long-form recording and streaming scenarios.

\section*{Limitations}
\textbf{Data Scale and Generalizability:} We trained and evaluated our models on AMI and AliMeeting, which represent challenging real-world conditions with spontaneous overlapping speech and noise. However, the training data size for both datasets is relatively small ($65$ and $103$ hours respectively). Validating our approach on large-scale synthetic datasets would be beneficial to better assess its generalizability and scaling behavior compared to baselines. Furthermore, expanding the data scale would allow us to investigate whether occasional failures, such as structural parsing errors or hallucinations, are intrinsic to the architecture or a symptom of data scarcity.

\textbf{Inference Latency:} As an LLM-based framework, our model relies on autoregressive generation, which is inherently slower than non-autoregressive discriminative models. Although we have verified that long-form recording (e.g., 30--50 minutes) can be decoded via chunk-wise inference by adopting a "best-of-both-worlds" strategy~\citep{kinoshita2021integrating}, the overall inference speed still lags behind highly optimized cascaded systems. This latency currently constrains the model’s deployment in strictly real-time or low-resource streaming scenarios.

\textbf{Scope of Multimodal Understanding:} Ideally, a comprehensive multi-speaker interaction model should capture not only identity and content but also paralinguistic attributes such as gender, age, and emotional state. The speaker encoder we use, Auden-Voice, possesses pre-trained capabilities for these tasks, and our preliminary experiments suggested promising predictive potential. However, since the AMI and AliMeeting corpora are business meetings and lack rigorous annotations for such paralinguistic information, we were unable to perform a quantitative evaluation in this study. Future work will explore datasets with richer paralinguistic labels to fully unlock this potential.

\section*{Acknowledgement}
This work used the Delta system at the National Center for Supercomputing Applications through allocation CIS250338 from the Advanced Cyberinfrastructure Coordination Ecosystem: Services \& Support (ACCESS) program \cite{boerner2023access}, which is supported by U.S. National Science Foundation grants \#2138259, \#2138286, \#2138307, \#2137603, and \#2138296. Yuheng Zhang is supported by a fellowship from the Amazon-Illinois Center on AI for Interactive Conversational Experiences (AICE).

We would like to thank the area chair and the anonymous reviewers for their constructive feedback, which greatly helped improve the clarity of our methodology and the depth of our experimental analyses, particularly regarding multi-speaker overlap handling and cross-lingual evaluation.

\bibliography{ref}

\appendix

\section{Appendix}
\label{sec:appendix}

\subsection{Dataset and Metric Details}
\label{sec:sec:dataset}

Table~\ref{tab:dataset} summarizes the statistics of two datasets of total hours and number of utterances. Since all compared models show unstable behavior on extreme-duration inputs, we then filter out utterances shorter than 0.5 seconds (mainly filler words such as “okay”) and longer than 80 seconds during evaluation. This removes approximately 5\% of total speech duration from each dataset to ensure a fair comparison.

Both AMI and AliMeeting datasets use anonymized, code-based speaker identifiers instead of real names.
\begin{table}[H]
    \centering
    \resizebox{\linewidth}{!}{
    \begin{tabular}{l|cc|cc}
        \toprule
        \multirow{2}{*}{\textbf{Split}} & \multicolumn{2}{c|}{\textbf{AMI}} & \multicolumn{2}{c}{\textbf{AliMeeting}} \\
        & Hours & \#Samples & Hours & \#Samples \\
        \midrule
        Train        & 65.0 & 26,760 & 103.4 & 38,338 \\
        Dev./Eval.   &  7.6 &  3,887 &   3.9 &  1,703 \\
        Test         &  7.3 &  3,065 &   9.9 &  4,585 \\
        \bottomrule
    \end{tabular}
    }
    \caption{Statistics of training, validation, and test splits for AMI-SDM and AliMeeting-Far datasets. Both are processed with the Lhotse toolkit~\cite{Zelasko_Lhotse_a_speech_2021}}
    \label{tab:dataset}
\end{table}

\paragraph{Diarization Error Rate (DER)} is computed as:
\begin{equation*}
\text{DER} = \frac{T_{\text{FA}} + T_{\text{MISS}} + T_{\text{CONF}}}{T_{\text{TOTAL}}}
\end{equation*}
where $T_{\text{FA}}$, $T_{\text{MISS}}$, $T_{\text{CONF}}$ denote the durations of false alarm speech (predicted speakers exceed reference speakers), missed speech (predicted speakers are fewer than reference speakers), and speaker confusion (correct speaker count but incorrect speaker assignment), respectively. We use a 0~ms collar with no forgiveness for overlapping speech.

\paragraph{Word Error Rate (WER)} is defined as:
\begin{equation*}
\text{WER} = \frac{S + D + I}{N}
\end{equation*}
where $S$, $D$, $I$ are the number of substitutions, deletions, and insertions, respectively. $N$ is total number of words in the reference. Mandarin uses Character Error Rate (CER).

\paragraph{Global WER (gWER)} is calculated as the WER over all utterances concatenated chronologically, regardless of speaker labels.

\paragraph{Concatenated Minimum Permutation WER (cpWER)} is defined as:
\begin{equation*}
\text{cpWER} = \min_{\pi} \frac{\sum_i \text{WER}(R_i, H_{\pi(i)}) \cdot |R_i|}{\sum_i |R_i|}
\end{equation*}
where $R_i$ and $H_j$ are the reference and predicted transcripts for speaker $i$ and $j$, respectively, and $\pi$ denotes the optimal assignment via Hungarian algorithm.

For reproducibility, DER and cpWER are computed using the MeetEval Toolkit~\citep{von2023meeteval}.

\subsection{Reproducibility Details}
\label{sec:sec:reproduce}

\paragraph{Acoustic Features.}
We extract 80-dimensional log-mel spectrogram features with a window size of 25 ms and hop size of 10 ms.

\paragraph{Encoder Architecture.} Both encoders share the same backbone: Zipformer. The output embeddings are downsampled to 25 Hz, with a hidden dimension $D_{enc}=768$. Each encoder has approximately 156M parameters. Each encoder is followed by a projector with a stride $k=4$, yielding compressed representations at 6.25 Hz. Both projectors are trained from scratch, trainable parameters are 47.72M in total.

\paragraph{Time Anchor Mechanism.}
We set the numeric time anchor interval $m=8$, corresponding to 1.28 physical seconds given the 6.25 Hz compressed frame rate. Since Qwen2.5's tokenizer splits multi-digit numbers into separate digit tokens (e.g., \texttt{12} becomes \texttt{1, 2}), we pre-cache the embeddings for digits \texttt{0–9} to facilitate training.

\paragraph{Large Language Model.}
The LLM backbone is {Qwen2.5-7B-Instruct} with $D_{LLM}=3584$. It is queried with \texttt{use\_flash\_attn}, and used in inference-only mode. Greedy decoding is used throughout. We use its default tokenizer and vocabulary, with an additional special token \texttt{<|AUDIO|>} as placeholder for actual features sequences.

\paragraph{Data Sampling and Augmentation.}
We use a bucketing sampler with a maximum input duration of 80 seconds as training batch size. Utterances are shuffled during training. SpecAugment is used for data augmentation. Speed perturbation is disabled.

\paragraph{Training Setup.}
The model is trained using the Adam optimizer with a learning rate of 0.001. Learning rate scheduling is handled by the \texttt{eden} scheduler, configured with \texttt{lr\_epochs} = 3.5, \texttt{lr\_batches} = 7500, \texttt{warmup\_batches} = 500, and \texttt{lr\_steps\_per\_epoch} = 0. Mixed-precision training is enabled. Decoding is performed via greedy search. Flash attention is disabled.

\paragraph{Hardware.}
All training and evaluation experiments are conducted on a single NVIDIA H200 or A100 GPU. On an H200, the model converges around 20k steps in approximately  3 hours; on an A100, takes around 8 hours.

\subsection{Baseline}
\label{sec:sec:baseline}

\paragraph{Gemini}
We use \texttt{Gemini-2.0-flash} via the official \texttt{genai} API to perform end-to-end speaker diarization and multi-speaker ASR. All \texttt{safety\_settings} are disabled to allow unrestricted decoding. We adopt a structured prompt to encourage overlapping utterance segmentation. However, the model exhibits a systematic limitation: it only generates segments in \textbf{strictly sequential order} (e.g., 1.5–2.0, 2.1–3.0) without any temporal overlap, even when overlapping speech is clearly present in the audio. This behavior persists despite explicit prompting, and hinders the model's ability for precise DER.

Following Google’s specification\footnote{\url{https://ai.google.dev/api/generate-content\#FinishReason}}, if the returned \texttt{FinishReason} equals 2 (\texttt{MAX\_TOKENS}), the output is often hallucinated or incomplete; we mark such cases as failure. We also evaluated more recent \texttt{Gemini-2.5-pro} and \texttt{Gemini-3.0-pro-preview}, but they suffered much more frequent truncation errors (around 30\% for the first 100 samples), so we only report results from \texttt{Gemini-2.0-flash}.

For Mandarin outputs, the model occasionally returns Traditional Chinese. We use \texttt{OpenCC}\footnote{\url{https://github.com/BYVoid/OpenCC}} to convert all transcripts into Simplified Chinese before computing WER.

\begin{tcblisting}{
  title=\textbf{Prompt for Gemini-2.0-flash},
  listing only,
  listing options={
    basicstyle=\ttfamily\small,
    breaklines=true,
    columns=fullflexible
  }
}
Perform speaker diarization and ASR on English (when decode Alimeeting, "on Mandarin Chinese"). It may contain OVERLAPPING speech, make each utterance turn complete, do NOT chop segments to avoid overlap. If no speech detected in a segment, skip it

Output a JSON list:
[
  {
    "start": 0.0,
    "end": 1.0,
    "speaker": "spk_1",
    "text": "..."
  }
]

Decoding Configuration:
- Temperature: 0.0
- Maximum output tokens: 2048
- Response mime type: application/json
\end{tcblisting}

\paragraph{Qwen-Omni Models} We evaluate two end-to-end models from the Qwen Omni family: \texttt{Qwen2.5-Omni-7B}\footnote{\url{https://huggingface.co/Qwen/Qwen2.5-Omni-7B}} and \texttt{Qwen3-Omni-30B-A3B-Instruct}\footnote{ \url{https://huggingface.co/Qwen/Qwen3-Omni-30B-A3B-Instruct}}. These models are capable of generating both text and audio outputs. Since our task only requires textual diarization and ASR results, we explicitly disable audio generation. Similar to our observations with Gemini, the Qwen models also output speaker segments in strictly linear timelines. Furthermore, \texttt{Qwen3-Omni} adopts a SubRip Subtitle (SRT)-like format (e.g., \texttt{"00:00.000"}), which we normalize during post-processing to ensure consistent evaluation.

\begin{tcblisting}{
  title=\textbf{Prompt for Qwen2.5/3-Omni},
  colback=blue!3,
  colframe=blue!45,
  colbacktitle=blue!15,
  coltitle=black,
  % fonttitle=\small,
  % arc=0mm,
  % boxrule=0.5pt,
  listing only,
  listing options={
    basicstyle=\ttfamily\small,
    breaklines=true,
    columns=fullflexible,
    showstringspaces=false
  }
}
System Message:
You are a speech analysis engine. Output ONLY JSON.

User Prompt:
Task: Speaker Diarization & ASR (English for AMI/ Mandarin for Alimeeting).
Instructions:
1. Output a JSON list of segments.
2. Keys: "start", "end", "speaker", "gender", "text".
3. It may contain OVERLAPPING speech, make each utterance turn complete, do NOT chop segments to avoid overlap. 
4. No preamble. No explanations.

Decoding Configuration:
- Disable talker: True
- Sampling: disabled
- Maximum new tokens: 2048
\end{tcblisting}

\paragraph{Pyannote} As part of our cascade pipeline, we use the Pyannote pipeline\footnote{\url{https://huggingface.co/pyannote/speaker-diarization-3.1}}. This model is widely adopted, optimized specifically for diarization, and offers strong performance with fast inference. However, for short utterances (e.g., $\sim$1 second), especially backchanneling, the model often fails to detect any speaker activity and outputs nothing, thus affecting downstream speech recognition handled by Whisper.

\paragraph{Whisper}
For the second stage of the cascade pipeline, we use \texttt{whisper-large-v3}\footnote{\url{https://huggingface.co/openai/whisper-large-v3}}, a widely adopted strong baseline for multi-lingual speech recognition. We apply it to each segment predicted by Pyannote. The decoding language is explicitly set: \texttt{en} for English and \texttt{zh} for Mandarin. For Mandarin segments, the model often produces a mix of Simplified and Traditional Chinese characters. We also post-process the transcripts using \texttt{OpenCC} before computing WER.

\paragraph{WavLM}

In Section~\ref{sec:ablation_sot}, we compare our dual-encoder architecture against \texttt{WavLM-Large}\footnote{\url{https://huggingface.co/microsoft/wavlm-large}}, designed as a general-purpose speech representation model ``for full-stack speech processing.'' The model contains 316.62M parameters, comparable to the combined size of our dual encoders.

\subsection{Cross-dataset and Cross-lingual Generalization}
\label{sec:sec:appendix_cross_lingual}
To evaluate the generalization capability of TagSpeech across different languages and acoustic environments, we conducted zero-shot cross-dataset experiments. We tested the model trained on AMI (English) directly on the AliMeeting (Mandarin) test set, and vice versa. 

The results, summarized in Table~\ref{tab:cross_lingual}, reveal a compelling decoupling between temporal grounding and linguistic decoding. Despite the extreme language shift, DER remains competitive on successfully parsed samples. This confirms that the time-anchor and SOT mechanisms in TagSpeech learn \textbf{language-agnostic} representations for multi-speaker temporal grounding.

\begin{table}[ht]
\centering
\small
\setlength{\tabcolsep}{4pt}
\begin{tabularx}{\columnwidth}{l @{\extracolsep{\fill}} ccc}
\toprule
\textbf{Train $\rightarrow$ Test} & \textbf{Fail $\downarrow$} & \textbf{DER $\downarrow$} & \textbf{cpWER $\downarrow$} \\ \midrule
AMI (EN) $\rightarrow$ AliMeeting (ZH) & 28.20 & 19.99 & $>100$ \\
AliMeeting (ZH) $\rightarrow$ AMI (EN) & 22.94 & 26.53 & $>100$ \\ \bottomrule
\end{tabularx}
\caption{Zero-shot cross-lingual evaluation between English (AMI) and Mandarin (AliMeeting).}
\label{tab:cross_lingual}
\end{table}

However, the cpWER exceeds 100\% in both cross-lingual settings. Qualitative analysis reveals a strong language-specific hallucination: the model generates hypotheses exclusively in the training language, regardless of the input audio. This suggests that while the frozen LLM backbone maintains the SOT format, its internal language modeling is heavily biased, causing content recognition to collapse under extreme language shift. These findings further motivate our future plan to apply LoRA-based adaptation to the LLM backbone, aiming to enhance linguistic generalization while preserving robust temporal grounding.

\subsection{Full Results}
\label{sec:sec:full_results}

We report the main results on the standard test sets of AMI and AliMeeting. Due to the high computational cost of training and inference with LLM backbones, we report results from a single run with a fixed random seed to ensure reproducibility. The full results corresponding to Table~\ref{tab:main_results} are presented here in Table~\ref{tab:full_results}. We highlight catastrophic outliers in red and summarize several descriptive observations below:

\begin{table*}[t]
\scriptsize
\centering
\begin{adjustbox}{max width=\textwidth}
\begin{tabularx}{\textwidth}{l|l|
    >{\centering\arraybackslash\hsize=0.85\hsize}X 
    >{\centering\arraybackslash\hsize=1.6\hsize}X  
    >{\centering\arraybackslash\hsize=0.85\hsize}X 
    >{\centering\arraybackslash\hsize=0.85\hsize}X 
    >{\centering\arraybackslash\hsize=0.85\hsize}X 
    |
    >{\centering\arraybackslash\hsize=0.85\hsize}X 
    >{\centering\arraybackslash\hsize=1.6\hsize}X  
    >{\centering\arraybackslash\hsize=0.85\hsize}X 
    >{\centering\arraybackslash\hsize=0.85\hsize}X 
    >{\centering\arraybackslash\hsize=0.85\hsize}X 
}
\toprule
\textbf{Model} & \textbf{Group} 
& \multicolumn{5}{c|}{\textbf{AMI-SDM (English)}} 
& \multicolumn{5}{c}{\textbf{AliMeeting (Mandarin)}} \\
\cmidrule(lr){3-7} \cmidrule(lr){8-12}
& & Fail ↓ & DER ↓ 0.0|0.25 & cpWER↓ & gWER ↓ & SCA ↑
  & Fail ↓ & DER ↓ 0.0|0.25 & cpWER↓ & gWER ↓ & SCA ↑ \\
\midrule
\multirow{8}{*}{Gemini-2.0-flash}
& 0.5–1s   & 0.9 & 32.16 | 28.73 & \cellcolor{red!25}{260.63} & \cellcolor{red!25}{260.08} & 79.82
           & \cellcolor{red!25}{36.2} & 75.20 | 73.11 & \cellcolor{red!25}{2141.57} & \cellcolor{red!25}{2140.43} & 24.40 \\
& 1–15s    & 0.5 & 44.15 | 36.14 & 36.33 & 32.16 & 57.19
           & 32.7 & 43.96 | 35.18 & 43.34 & 38.78 & 60.27 \\
& 15–30s   & 0.9 & 47.06 | 41.39 & 31.92 & 25.27 & 40.23
           & 20.3 & 55.04 | 47.95 & 49.82 & 35.24 & 21.26 \\
& >30s     & \cellcolor{red!25}{23.7} & 44.87 | 39.03 & 32.82 & 22.83 & 26.52
           & \cellcolor{red!25}{35.5} & 57.63 | 51.69 & 59.14 & 37.71 & 13.51 \\
\cmidrule(lr){2-12}
& 1 spk    & 0.5 & 32.19 | 29.92 & 28.61 & 28.44 & 96.39
           & \cellcolor{red!25}{40.8} & 33.26 | 28.48 & 89.90 & 89.75 & 83.24 \\
& 2 spk    & 0.7 & 41.33 | 35.07 & 28.28 & 26.31 & 27.94
           & 29.6 & 43.39 | 36.38 & 33.95 & 29.55 & 38.16 \\
& 3 spk    & 3.5 & 51.55 | 44.80 & 40.48 & 31.51 & 3.86
           & 0.6 & 49.05 | 42.45 & 45.04 & 36.48 & \cellcolor{red!25}{0.29} \\
& 4 spk    & 13.6 & 57.32 | 50.16 & 49.75 & 33.87 & \cellcolor{red!25}{0.54}
           & 11.0 & 63.83 | 57.87 & 66.52 & 44.67 & \cellcolor{red!25}{0.28} \\

\midrule
\midrule
\multirow{8}{*}{\parbox{2cm}{\centering Qwen2.5-Omni\\-7B}}
& 0.5–1s   & 1.2 & 6.60 | 1.48 & 59.73 & 58.22 & 87.73
           & 0.8 & 6.61 | 1.85 & 25.53 & 25.03 & 90.75 \\
& 1–15s    & 3.3 & 26.62 | 20.39 & 45.06 & 34.21 & 62.36
           & 6.0 & 24.99 | 19.51 & 26.96 & 19.02 & 74.30 \\
& 15–30s   & 4.6 & 32.21 | 26.35 & 47.92 & 30.96 & 39.10
           & 7.6 & 46.03 | 26.71 & 48.07 & 26.71 & 40.83 \\
& >30s     & 16.2 & 49.91 | 45.43 & 58.05 & 30.03 & 19.31
           & 24.4 & 57.81 | 34.67 & 67.47 & 34.67 & 24.62 \\
\cmidrule(lr){2-12}
& 1 spk    & 3.0 & 8.64 | 7.75 & 26.42 & 23.25 & 92.18
           & 6.2 & 11.09 | 10.50 & 8.77 & 7.78 & 93.28 \\
& 2 spk    & 3.2 & 30.00 | 25.90 & 40.80 & 30.36 & 53.70
           & 3.9 & 32.35 | 28.73 & 30.03 & 16.56 & 66.64 \\
& 3 spk    & 4.4 & 44.58 | 40.16 & 60.13 & 35.66 & \cellcolor{red!25}{0.73}
           & 6.1 & 44.34 | 39.87 & 50.70 & 29.10 & 1.24 \\
& 4 spk    & 12.2 & 54.66 | 49.96 & 70.10 & 38.93 & \cellcolor{red!25}{0.00}
           & 15.5 & 59.73 | 54.98 & 72.28 & 42.34 & 2.95 \\
\midrule
\midrule
\multirow{8}{*}{\parbox{2cm}{\centering Qwen3-Omni\\-30B-A3B-Instruct}}
& 0.5–1s   & 13.9 & 11.58 | 2.11 & 51.78 & 51.19 & 88.03
           & 0.5 & 9.20 | 1.04 & 19.22 & 18.62 & 93.85 \\
& 1–15s    & 0.2 & 31.31 | 23.23 & 33.37 & 28.33 & 61.49
           & 0.2 & 21.39 | 13.81 & 19.79 & 15.92 & 73.17 \\
& 15–30s   & 0.3 & 32.84 | 25.51 & 29.87 & 23.60 & 42.29
           & 0.5 & 34.32 | 25.31 & 33.39 & 24.21 & 41.49 \\
& >30s     & 3.5 & 55.23 | 49.03 & 52.77 & 48.15 & 23.95
           & 4.1 & 61.08 | 53.61 & 61.15 & 54.40 & 26.06 \\
\cmidrule(lr){2-12}
& 1 spk    & 3.6 & 19.42 | 18.38 & 20.89 & 18.82 & 93.92
           & 0.2 & 8.91 | 7.19 & 7.24 & 5.68 & 94.75 \\
& 2 spk    & 0.6 & 31.17 | 25.17 & 30.38 & 27.31 & 48.02
           & 0.1 & 23.71 | 16.92 & 18.44 & 15.83 & 60.24 \\
& 3 spk    & 0.2 & 46.34 | 38.83 & 45.65 & 39.00 & 4.90
           & 0.9 & 36.56 | 29.67 & 36.23 & 29.59 & 2.94 \\
& 4 spk    & 2.3 & 56.40 | 49.19 & 54.78 & 45.86 & 3.85
           & 2.2 & 59.19 | 51.79 & 61.42 & 50.26 & 6.12 \\
\midrule
\midrule
\multirow{8}{*}{\parbox{2cm}{\centering Cascade System:\\Pyannote-3.1+\\Whisper-large-v3}}
& 0.5–1s   & \cellcolor{red!25}{21.8} & 21.41 | 7.15 & 53.47 & 51.82 & 86.43
           & 19.6 & 15.22 | 4.00 & 45.03 & 45.03 & 94.60 \\
& 1–15s    & 2.7 & 21.24 | 13.20 & 42.43 & 35.81 & 51.23
           & 1.0 & 17.52 | 9.46 & 36.58 & 30.80 & 62.20 \\
& 15–30s   & 0.0 & 21.79 | 14.83 & 40.96 & 33.34 & 38.75
           & 0.0 & 28.08 | 18.34 & 47.66 & 40.28 & 31.98 \\
& >30s     & 0.6 & 26.37 | 17.39 & 46.84 & 37.97 & 26.74
           & 0.0 & 40.06 | 30.20 & 63.68 & 54.68 & 38.95 \\
\cmidrule(lr){2-12}
& 1 spk    & 8.4 & 6.06 | 4.96 & 26.12 & 25.51 & 99.40
           & 4.0 & 2.63 | 0.89 & 17.80 & 17.63 & 99.79 \\
& 2 spk    & 1.5 & 17.81 | 11.53 & 39.55 & 33.60 & 10.37
           & 1.0 & 17.17 | 9.66 & 35.97 & 29.65 & 14.97 \\
& 3 spk    & 0.0 & 28.10 | 19.08 & 48.94 & 38.99 & 7.44
           & 0.3 & 27.88 | 19.13 & 55.82 & 46.97 & 7.02 \\
& 4 spk    & 0.9 & 37.17 | 26.89 & 55.64 & 43.33 & 6.16
           & 0.0 & 46.84 | 37.42 & 69.01 & 57.79 & 11.97 \\
\midrule
\midrule
\multirow{8}{*}{TagSpeech (Ours)}
& 0.5–1s   & 0.3 & 8.25 | 0.62 & 45.45 & 43.83 & 90.27
           & 0.0 & 4.46 | 1.22 & 24.25 & 23.95 & 93.88 \\
& 1–15s    & 0.2 & 16.26 | 8.90 & 35.74 & 28.48 & 72.04
           & 0.6 & 13.22 | 7.47 & 23.42 & 19.49 & 82.51 \\
& 15–30s   & 1.7 & 20.33 | 14.34 & 37.67 & 26.84 & 53.33
           & 7.6 & 28.83 | 21.86 & 42.01 & 29.72 & 66.93 \\
& >30s     & 12.7 & 42.61 | 37.40 & 56.75 & 40.36 & 40.40
           & \cellcolor{red!25}{45.9} & 47.12 | 42.57 & 61.16 & 41.60 & 59.14 \\
\cmidrule(lr){2-12}
& 1 spk    & 0.1 & 2.66 | 1.69 & 15.19 & 14.67 & 96.48
           & 0.0 & 1.11 | 0.69 & 8.45 & 8.22 & 97.09 \\
& 2 spk    & 0.4 & 18.16 | 12.98 & 34.38 & 26.42 & 61.29
           & 0.5 & 16.94 | 11.85 & 26.74 & 19.60 & 66.58 \\
& 3 spk    & 3.3 & 32.77 | 27.18 & 52.76 & 37.73 & 31.97
           & 5.2 & 30.68 | 24.92 & 45.67 & 32.35 & 42.77 \\
& 4 spk    & 7.0 & 45.54 | 40.57 & 66.02 & 46.59 & 9.09
           & 27.2 & 49.01 | 44.71 & 66.12 & 49.20 & 56.51 \\
\bottomrule
\end{tabularx}
\end{adjustbox}
\caption{Full results of our model TagSpeech and baselines on AMI-SDM and AliMeeting-Far , grouped by utterance duration and speaker count. We evaluate on utterances with durations in the range of 0.5–80 s, excluding approximately 5\% of samples with extreme durations in both datasets. Each row is a group; Columns report metrics in percentage (\%), with DER presented under both 0.0s and 0.25s collars for comprehensive comparison. Cells highlighted in red indicate catastrophic outliers.}
\label{tab:full_results}
\end{table*}

\paragraph{By utterance duration.}
Across models, the most challenging cases are extremely short utterances (0.5–1s) and long utterances (>30s). For Gemini-2.0-flash, extremely short segments exhibit exceptionally high WER, indicating severe failure behavior due to hallucination.

\paragraph{By speaker count.}
Performance consistently degrades as the number of speakers increases in the ground truth. This trend is observed across all metrics and all models. In particular, Qwen2.5 shows near-zero speaker count accuracy for utterances with more than two speakers, suggesting limited sensitivity to multi-speaker presence.

\paragraph{Stability across conditions.}
The cascade system shows relatively stable performance across different utterance durations and speaker counts, except for extremely short segments where failures often occur at the diarization stage. Overall metric variance with respect to duration is smaller compared to end-to-end LLM-based models. In addition, Qwen-3.0-Omni 30B model exhibits the lowest failure rate across most conditions.

\paragraph{DER Collar 0.25\,s}
While we prioritize the strict 0.0\,s collar in all tables to better reflect the fine-grained temporal boundary modeling of TagSpeech, we also provide results using a 0.25\,s collar to facilitate a broader comparison with existing literature and prior work. These additional results, summarized in Table~\ref{tab:collar_results}, demonstrate that performance trends remain consistent with the collar-free evaluation.

\begin{table}[H]
\centering
\small
\begin{tabularx}{\columnwidth}{l @{\extracolsep{\fill}} cc}
\toprule
\textbf{System} & \textbf{AMI-SDM} & \textbf{AliMeeting-Far} \\ \midrule
Gemini-2.0-flash & 38.64 & 42.50 \\
Qwen2.5-Omni-7B & 29.39 & 31.86 \\
Qwen3-Omni-30B & 31.94 & 25.93 \\
Pyannote + Whisper & 15.04 & 16.71 \\
TagSpeech (Ours) & 18.73 & 16.18 \\ \bottomrule
\end{tabularx}
\caption{Diarization Error Rate (DER \% $\downarrow$) with a standard 0.25\,s collar. These results correspond to the DER measurements in Table~\ref{tab:main_results} and facilitate a broader comparison with existing literature.}
\label{tab:collar_results}
\end{table}

\end{document}